\documentclass[letter]{aa}

\usepackage{txfonts}

\usepackage{graphicx}   
\usepackage{amsmath}    
\usepackage{amssymb}    
\usepackage{xcolor}
\usepackage[normalem]{ulem}
\usepackage{bm}
\usepackage[nomarkers,tablesonly]{endfloat}
\usepackage{hyperref}
\hypersetup{colorlinks,
        citecolor=[rgb]{0.1 , 0.3, 0.8}}

\newcommand{\lya}{\mbox{${\rm Ly}\alpha$}}

\newcommand{\zabs}{\ensuremath{z_{\rm abs}}}

\newcommand{\Rcnm}{\ensuremath{R_{\textsc{cnm}}}}

\newcommand{\CI}{\ion{C}{i}}

\newcommand{\HI}{\ion{H}{i}}
\newcommand{\NHI}{\ensuremath{N_{\rm H\, \textsc{i}}}}
\newcommand{\NHH}{\ensuremath{N_{\rm H_2}}}
\newcommand{\NHtot}{\ensuremath{N_{\rm H}}}
\newcommand{\Pmin}{\ensuremath{P_{\textrm{min}}}}

\newcommand{\logNHIcm}{\ensuremath{\log({\rm N_{H\, \textsc{i}}\: /\: cm}^{-2})}}
\newcommand{\lCNM}{\ensuremath{l_{\textsc{cnm}}}}
\newcommand{\lCI}{\ensuremath{l_{\textsc{Ci}}}}
\newcommand{\Iuv}{\ensuremath{I_{\textsc{uv}}}}

\defcitealias{Krogager2020}{K20}        
\defcitealias{Balashev2018}{BN18}

\begin{document}


\title{Modeling the statistics of the cold neutral medium in absorption-selected high-redshift galaxies}
\titlerunning{Modeling CNM statistics in high-$z$ quasar absorbers}

\author{
        Jens-Kristian Krogager\inst{1}
        \and
        Pasquier Noterdaeme\inst{1}
}

\institute{ Institut d'Astrophysique de Paris,
            CNRS-SU, UMR7095, 98bis bd Arago, 75014 Paris, France\\
            \email{krogager@iap.fr}
}

\abstract{We present a statistical model of the selection function of cold neutral gas in high-redshift ($z=2.5$) absorption systems. The model is based on the canonical two-phase model of the neutral gas in the interstellar medium and contains only one parameter for which we do not have direct observational priors: namely the central pressure of an $L^*$ halo at $z=2.5$, $P^*$. Using observations of the fraction of cold gas absorption in strong \HI-selected absorbers, we were able to constrain $P^*$. The model simultaneously reproduces the column density distributions of \HI\ and H$_2$, and we derived an expected total incidence of cold gas at $z\sim2.5$ of $\lCNM = 12\times 10^{-3}$. Compared to recent measurements of the incidence of \CI-selected absorbers (EW$_{\lambda\, 1560} > 0.4$~\AA), the value of \lCNM\ from our model indicates that only 15\% of the total cold gas would lead to strong \CI\ absorption (EW~$ > 0.4$~\AA). Nevertheless, \CI\ lines are extremely useful probes of the cold gas as they are relatively easy to detect and provide direct constraints on the physical conditions. 
Lastly, our model self-consistently reproduces the fraction of cold gas absorbers as a function of \NHI. 
}


\maketitle

\section{Introduction}
\label{intro}

    Our understanding of star formation throughout cosmic time is intimately linked to our ability to observe and constrain the physical properties of the gas in and around galaxies. The neutral gas is of particular interest and tends to split into a warm, diffuse ($T \sim 10^4$~K, $n\sim 0.5$\,cm$^{-3}$), and a cold, dense ($T \sim 100$~K, $n\sim 50$\,cm$^{-3}$) phase \citep*{Field1969}, the latter being more inclined to Jeans instability and subsequent star formation.

    \citet{Wolfire1995} described the two neutral phases as a result of the balance of heating and cooling mechanisms which, for a range in external pressure, exist in equilibrium.
    The minimum pressure required for a stable cold neutral medium (CNM) to exist, \Pmin, depends on several factors, out of which metallicity, $Z$, and ambient ionizing flux, \Iuv, play central roles.
    However, while the canonical two-phase description has been intensively investigated in nearby environments, these all have rather similar $Z$ and \Iuv. In order to test the current theoretical framework over an increased range of parameter space, it is convenient to look to high-redshift galaxies as the average \Iuv\ is higher \citep{Khaire2019} and the average metallicity is lower \citep[e.g.,][]{DeCia2018}.
    
    One powerful way of studying the neutral gas at high redshift is through damped \lya\ absorption systems (DLAs) observed in the spectra of distant quasars \citep[see review by][]{Wolfe2005}. 
    However, these high column density absorbers ($\NHI > 2 \times 10^{20}$~cm$^{-2}$) predominantly probe the warm and diffuse gas phase \citep[e.g.,][]{Srianand2012, Neeleman2015}.

    Instead, the cold gas phase can be studied directly by choosing appropriate tracers of the CNM (e.g., molecular hydrogen, H$_2$, neutral carbon, \ion{C}{i,} or absorption from \HI\ at 21-cm, since the 21-cm optical depth depends inversely on the temperature). Yet, such cold gas absorbers (hereafter referred to as CNM absorbers) are rare and only a few hundred -- compared to tens of thousands of DLAs -- have been identified in large-scale spectroscopic surveys \citep{Balashev2014, Ledoux2015, Srianand2012, Kanekar2014}. These CNM absorbers are extremely powerful probes of the physical conditions of the gas (density, temperature, and \Iuv) through atomic fine-structure and molecular transitions \citep[e.g.,][]{Srianand2005, Noterdaeme2007a}. It is thus possible to constrain the interstellar medium (ISM) conditions in galaxies out to high redshift $z\sim3$ \citep[e.g.,][]{Balashev2019}. Such direct observational constraints are crucial in order to test theoretical models and numerical simulations of star formation.
    
    Recent numerical simulations are able to follow the detailed evolution of H$_2$ directly and resolve the small-scale CNM (e.g., \citealt*{Nickerson2019}; \citealt{Bellomi2020}). However, the CNM cross section depends heavily on resolution and on the poorly constrained feedback mechanisms from star formation and supernovae. Nevertheless, the bulk of the CNM cross section arises in a centrally concentrated region with a uniform covering factor.

    In this letter, we model the CNM cross section at high redshift using a simple analytical description of the two-phase ISM. Our approach is based on the work by \citet[][hereafter K20]{Krogager2020} who present an effective model for DLAs building on the ideas of \citet{Fynbo2008}. Here we include a simple pressure-based prescription of the two-phase medium to model the total CNM incidence and the fraction of DLAs exhibiting CNM absorption.
    This way we are able to quantify the fraction of CNM absorbers that were selected using \CI\ absorption lines.
    
    Throughout this work, we assume a flat $\Lambda$CDM cosmology with $H_0=68\, \mathrm{km s}^{-1}\mathrm{Mpc}^{-1}$, $\Omega_{\Lambda}=0.69$, and $\Omega_{\textsc{m}} = 0.31$ \citep{Planck2016}.

\section{Literature data}
\label{data}

    We have compiled a sample of known CNM absorbers in the literature at $\zabs > 1.5$ with measurements of both metallicity and \NHI. The observable quantities used in this analysis are the column densities of neutral and molecular hydrogen, the gas-phase metallicity, and the thermal gas pressure. We furthermore included impact parameters for the seven absorption systems where an emission counterpart has been identified.
    The collected data are presented in Table~\ref{tab:data}. Only absorption systems with detections of \CI\ and/or H$_2$ were included in the compilation. The metallicities were calculated based on the elements Zn and S since these do not suffer from strong dust-depletion effects \citep{DeCia2016}.

\section{Modeling neutral gas absorption}
\label{model}

    The modeling of the neutral gas of high redshift galaxies followed the framework for DLA absorption by \citetalias{Krogager2020} and \citet{Fynbo2008}. Here we provide a brief summary of the model. Galaxies were drawn randomly from the UV luminosity function $\phi(L)$, which is described by a Schechter function: $\phi(L) = \phi_0 (L/L^*)^{\alpha} \exp{-L/L^*}$, with $\alpha=-1.7$. For each randomly drawn galaxy, we assigned an average projected extent of neutral gas ($R_{\textsc{dla}}$), a central metallicity ($Z_0$), and a radial metallicity gradient ($\gamma_Z$) based on empirically derived scaling relations, for details see \citetalias{Krogager2020}.
    An impact parameter ($b$) was assigned with a probability proportional to the area of a circular annulus at a given projected radius. The absorption metallicity was then calculated at the given impact parameter following the assumed metallicity gradient. Based on the absorption metallicity and the \NHI\ value along the line of sight (see Sect.~\ref{model:NHI}), we calculated the optical extinction ($A_V$) following \citet{Zafar2019}. We included a mock optical selection similar to large spectroscopic surveys by probabilistically rejecting sightlines with large $A_V$ \citep{Krogager2019}. For details regarding the implementation, we refer readers to \citetalias{Krogager2020}.

    In this work, we further model the radial pressure of the halo in order to calculate at what radial scales the ISM is able to support the CNM. We do this by assigning a halo mass to a given luminosity from which we can then calculate a halo pressure and its radial dependence. Since we are including a prescription for the pressure in this work, we are able to model \NHI\ more accurately than in our previous model \citepalias{Krogager2020}.
    The details of the pressure-based model are presented in what follows.

    \subsection{Modeling the cold neutral medium}

    Following \citet{Elmegreen1994}, we assume that $P_{\rm tot} \propto \Sigma_{\star}^2$. The radial dependence of the stellar surface density $\Sigma_{\star}$ then leads to a radial dependence on the pressure of the form:
    \begin{equation}
        \label{eq:Ptot}
            P_{\rm tot}(r) = P_0 \, e^{-2\,r/r_e}~.
    \end{equation}
    
    \noindent The effective radius $r_e$ scales with luminosity as:
    \begin{equation}
            \label{eq:re}
            r_e = r_e^* \, \left(\frac{L}{L^*}\right)^{t_e}~,
    \end{equation}
    \noindent where the characteristic scale length of an $L^*$ galaxy, $r_e^*$, is taken to be 3~kpc at $z=2.5$ \citep{vanderWel2014}, and the power-law index $t_e$ has a value of 0.3 \citep{Brooks2011}.
    The central pressure, $P_0$, is calculated as a function of halo mass, $M_h$, assuming that the central pressure scales with the virial pressure: $P_0 \propto T_{\rm vir}\, \rho_{\rm vir} \propto M_h^{2/3}$.
    The halo mass is assigned based on the luminosity according to the luminosity--$M_h$ relation by \citet{Mason2015}. The central pressure of a halo is then calculated as:
    \begin{equation}
        P_0 = P_0^* \, \left(\frac{M_h}{M_{h}^*}\right)^{2/3}~,
    \end{equation}
    \noindent
    where $M_h^* = 5\times 10^{12}$~M$_{\odot}$ at $z=2-3$ \citep{Mason2015}. The central pressure of an $L^*$ galaxy, $P_0^*$, is not constrained directly by observations. Instead, we explore a range of $10^5 - 10^7$~K~cm$^{-3}$.

    We approximate \Pmin\ following eq. (33) of \citet{Wolfire2003}:
    \begin{equation}
            \Pmin = \Pmin^* \, I_{\rm uv}\frac{Z_d / Z}{1 + 3.1 \left(\frac{I_{\rm uv} Z_d}{\zeta_{\rm cr}}\right)^{0.365}}~,
    \end{equation}
    \noindent
    where $I_{\rm uv}$ is the ambient UV field in units of the Draine field \citep{Draine1978}, $Z_d$ is the dust abundance, and $\zeta_{\rm cr}$ is the cosmic ray ionization rate.  In the above expression, $\Pmin^*$ refers to the minimum pressure at Solar metallicity with $I_{\rm uv} = 1$. We assume a fiducial value of $\Pmin^* = 10^4$~K~cm$^{-3}$ \citep{Wolfire2003}.
    We further assume that $I_{\rm uv} / \zeta_{\rm cr}$ is constant as both the cosmic ray ionization rate and the ambient UV field depend on the star formation activity. We model $Z_d/Z$ as a broken power-law following \citet[][their eq. 5]{Bialy2019}.
    
    The \Iuv\ is calculated based on the star formation rate surface density:
    \begin{equation}
        \label{eq:Iuv}
            I_{\rm uv} \propto \Sigma_{\textsc{sfr}} \propto \frac{\psi_{\star}}{r_e^2}~.
    \end{equation}

    \noindent The star formation rate, $\psi_{\star}$, is taken to be directly proportional to the UV luminosity, whereas the disk scale length is given above in Eq.~\eqref{eq:re}.

    The total \Iuv\ is given as the sum of the extragalactic UV background and the UV field related to on-going star formation. Hence, combining equations \eqref{eq:Iuv} and \eqref{eq:re}, assuming that the ambient UV field of an $L^*$ galaxy is one in units of the Draine field, results in an effective scaling between $I_{\rm uv}$ and $L$ of the form:
    \begin{equation}
            I_{\rm uv} = I_0 + \left(\frac{L}{L^*}\right)^{1 - 2t_e} = I_0 + \left(\frac{L}{L^*}\right)^{0.4}~,
    \end{equation}
    \noindent where $I_0 = 0.16$ is the extragalactic UV background integrated over $6-13.6$~eV at $z=2.5$ calculated by \citet{Khaire2019} in units of the Draine field.

    The radial dependence of the total pressure leads to a characteristic radial scale, \Rcnm, within which the total pressure is greater than \Pmin, and hence the ISM is able to support a stable CNM. This radial scale is determined as:
    \begin{equation}
            \Rcnm = \frac{1}{2}\, r_e\, \ln\left(\frac{P_0}{\Pmin}\right)~.
    \end{equation}

    For the ensemble of random sightlines drawn in the model described above, we then refer to sightlines as CNM absorbers if the impact parameter is less than \Rcnm, that is to say we implicitly assume a covering factor of one for the CNM for $r < \Rcnm$. The validity of this assumption is discussed in Sect.~\ref{discussion}.

    \subsection{Atomic and molecular hydrogen}
    \label{model:NHI}
    
    Given the inclusion of the halo pressure as a function of radius, we were able to model the neutral hydrogen in more detail than the average radial profile assumed by \citetalias{Krogager2020}. We started out by modeling the total hydrogen column, \NHtot, as a function of radius. Motivated by the universality of the exponential form as observed by \citet{Bigiel2012}, we used a central value of $\log \NHtot(r=0) = 21.5$ in units of cm$^{-2}$ for all galaxies. The exponential scale length was then matched to reproduce $\log \NHI = 20.3$ at the radius $R_{\textsc{dla}}$, which was set to match the observed incidence of DLAs \citepalias[see][]{Krogager2020}. In doing so, we assumed that \NHtot\ is dominated by \HI\ at these large radii, which is consistent with \citet{Bigiel2012}.
        
    We subsequently split the total hydrogen column density into separate atomic and molecular column densities using the pressure-based prescription by \citet{Blitz2006}:
    \begin{equation}
        f_{H_2} = \frac{N_{\rm H_2}}{\NHI} = \left(\frac{P_{\rm tot}}{P'}\right)^{0.92}~,
    \end{equation}
    \noindent with $P' = 4 \times 10^4$~K~cm$^{-3}$. Here, $P_{\rm tot}$ was calculated at the position of the random impact parameter following Eq.~\eqref{eq:Ptot}.
        
    In order to properly reproduce the distribution function of \NHI, we included a random scatter on \NHtot\ of 0.4~dex, mimicking the scatter in the observed radial profiles \citep{Bigiel2012}. The molecular fraction of a given absorber was furthermore given a random scatter of 0.1~dex according to the observations \citep{Bigiel2012}.
    We verified that this modeling of \NHI\ and $N_{\rm H_2}$ reproduces the distribution function of \NHI. The agreement is only slightly worse than the fit by \citetalias{Krogager2020}.

\section{Results}
\label{results}

    Based on a stacking experiment performed on SDSS spectra, \citet[hereafter BN18]{Balashev2018} infer the average covering fraction of self-shielded H$_2$-bearing gas among \HI-selected absorption systems of $4.0\pm0.5_{\rm stat}\pm1.0_{\rm sys}$~\% and $37\pm5_{\rm stat} \pm4_{\rm sys}$~\%, for $\logNHIcm > 20.3$ and $>21.7$, respectively.
	Here, we consider this molecular covering fraction as a proxy for the CNM fraction, that is to say the fraction of sightlines showing CNM absorption.
    
    The only a priori unknown parameter in our CNM model, $P_{0}^*$, was constrained by matching the CNM fractions by \citetalias{Balashev2018} using $\chi^2$-minimization. For this purpose, we used an effective uncertainty on the observed CNM fractions, combining the systematic and statistical uncertainties in quadrature. The best-fit value is $\log (P_{0}^* / k_B\, /\, {\rm K\, cm}^{-3}) = {5.95\pm0.11}$, for which we find the following average CNM fractions of 4.5\% and 34.8\% for $\logNHIcm > 20.3$ and $>21.7$, respectively. 
    This value of $P_{0}^*$ is consistent with the central pressure inferred for the Milky Way of $\log (P_0 / {\rm K\,cm^{-3}}) \approx 6$ given the local pressure measured in the Solar neighborhood of $\log (P_{r=R_{\odot}} / {\rm K\,cm^{-3}}) = 3.6$ \citep{Jenkins2011}, assuming $R_{\odot} = 8$~kpc and $r_{\rm h} = 3$~kpc.

    In order to check the resulting model distribution of thermal pressures, we compared our model to the observed thermal pressures, see Table~\ref{tab:data}. The observed average thermal pressure for CNM absorbers is $\langle \log(P_{\rm th} / {\rm K\, cm}^{-3}) \rangle = 4.0 \pm 0.1$, and our model predicts a value of $\langle \log(P_{\rm th} / {\rm K\, cm}^{-3}) \rangle = {3.8}$. Taking into account the caveat that the observations are neither homogeneous nor representative, the average pressure predicted by our model agrees well with the observations.

\begin{figure}
	\includegraphics[width=0.49\textwidth]{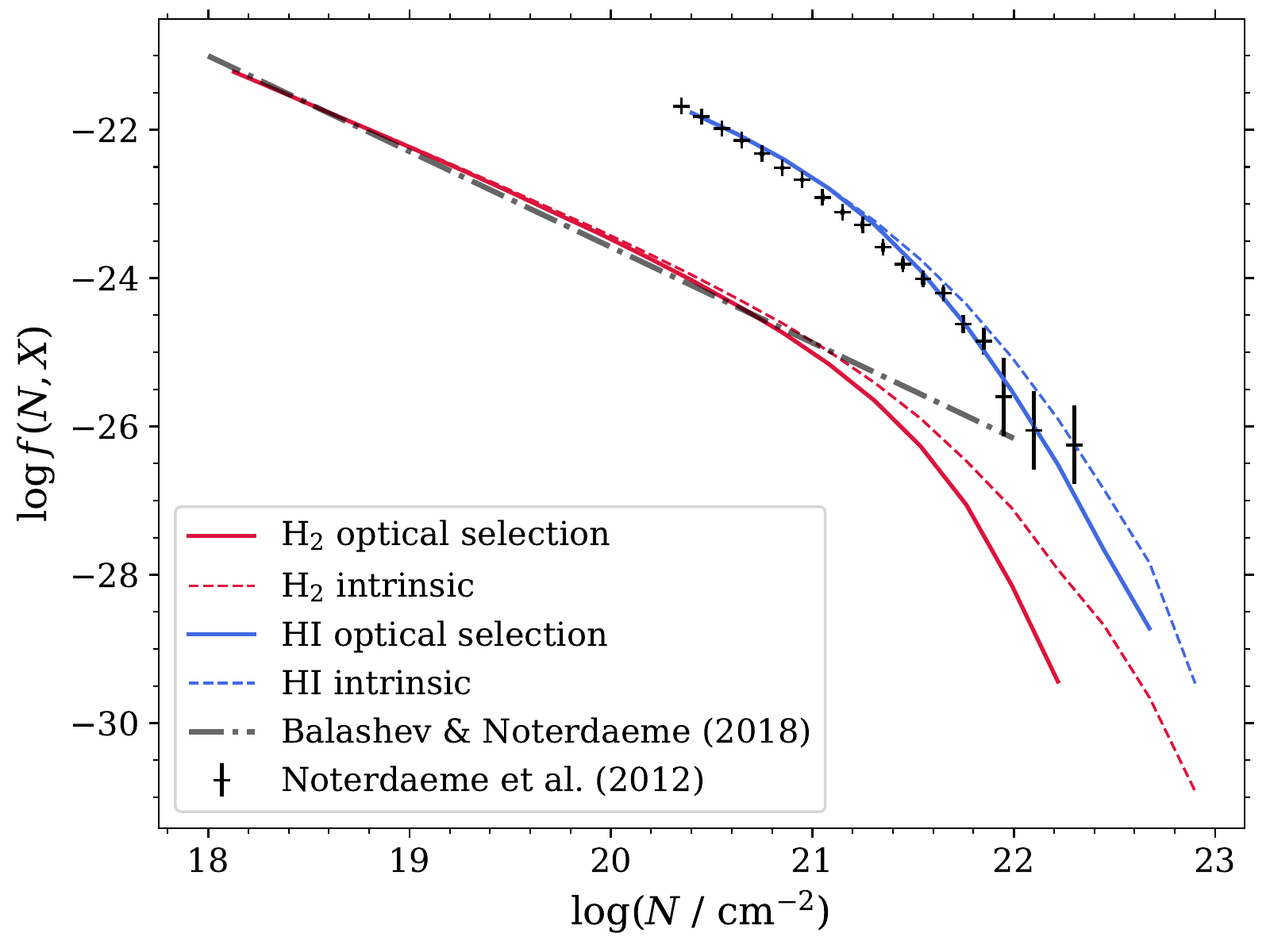}
    \caption{Predicted column density distribution functions of \NHI\ (blue) and \NHH\ (red) at $z\sim2.5$. The dashed lines indicate the intrinsic distributions and the solid lines are the "observable" distribution in optically-selected samples similar to SDSS-DR7.
    }
    \label{fig:fNH}
\end{figure}

    The predicted column density distribution functions $f(N, X)$ of \HI\ and H$_2$ are presented in Fig.~\ref{fig:fNH}. Overall, our model simultaneously reproduces the observed statistics of \NHI\ and \NHH. The power-law inferred from the stacking experiment by \citetalias{Balashev2018} represents the average shape of $f(N, X)$ and it is not sensitive to the "knee" at higher column densities. The disagreement at log(\NHH)$~\sim 22$ is therefore not surprising. We note that the location of the knee depends on the adopted central value of $N_{\rm H}$, which is motivated by observations by \citet{Bigiel2012}.

    The expected incidence of CNM absorbers was calculated as the integral:
    \begin{equation}
       \lCNM = \frac{{\rm d}n}{{\rm d}z} = c\, (1+z)^2\, H^{-1}(z) \int_{L_{\rm min}}^{\infty} \sigma(L)\, \phi(L)\, {\rm d} L~,
    \end{equation}

    \noindent
    where we adopted $L_{\rm min}=10^{-4} L^*$ following \citetalias{Krogager2020}; $\sigma(L)$ denotes the luminosity dependent absorption cross section, and the Hubble parameter is given as:
    \begin{equation}
        H(z) = H_0 \sqrt{\Omega_{\textsc{m}} (1+z)^3 + \Omega_{\Lambda}}~.
    \end{equation}
    
    \noindent The CNM absorption cross section was calculated based on the luminosity-dependent \Rcnm\,: $\sigma(L) = \pi \Rcnm^2$, which yields an incidence of $\lCNM = 11.3\times 10^{-3}$ at $z=2.5$ taking the dust obscuration bias into account.
    
    The incidence of \CI-selected absorbers inferred by \citet{Ledoux2015} at similar redshifts is $\lCI = 1.5 \pm 0.5 \times 10^{-3}$, which was corrected for incompleteness due to the limiting equivalent width.
    In order to compare our predicted \lCNM\ with that inferred by \citet{Ledoux2015}, we restricted our calculation to model points with $\log (Z_{\rm abs}/Z_{\odot}) > -0.6$, which corresponds to the minimum metallicity in the \CI-selected sample by \citet{Ledoux2015}. This yields an expected \CI\ incidence of $\lCI^{\prime} = 4.8\times 10^{-3}$. 
    A comparison between the overall expected incidence of cold gas and that traced by \CI\ absorption is discussed in Sect.~\ref{discussion}.

\begin{figure*}
	\includegraphics[width=0.95\textwidth]{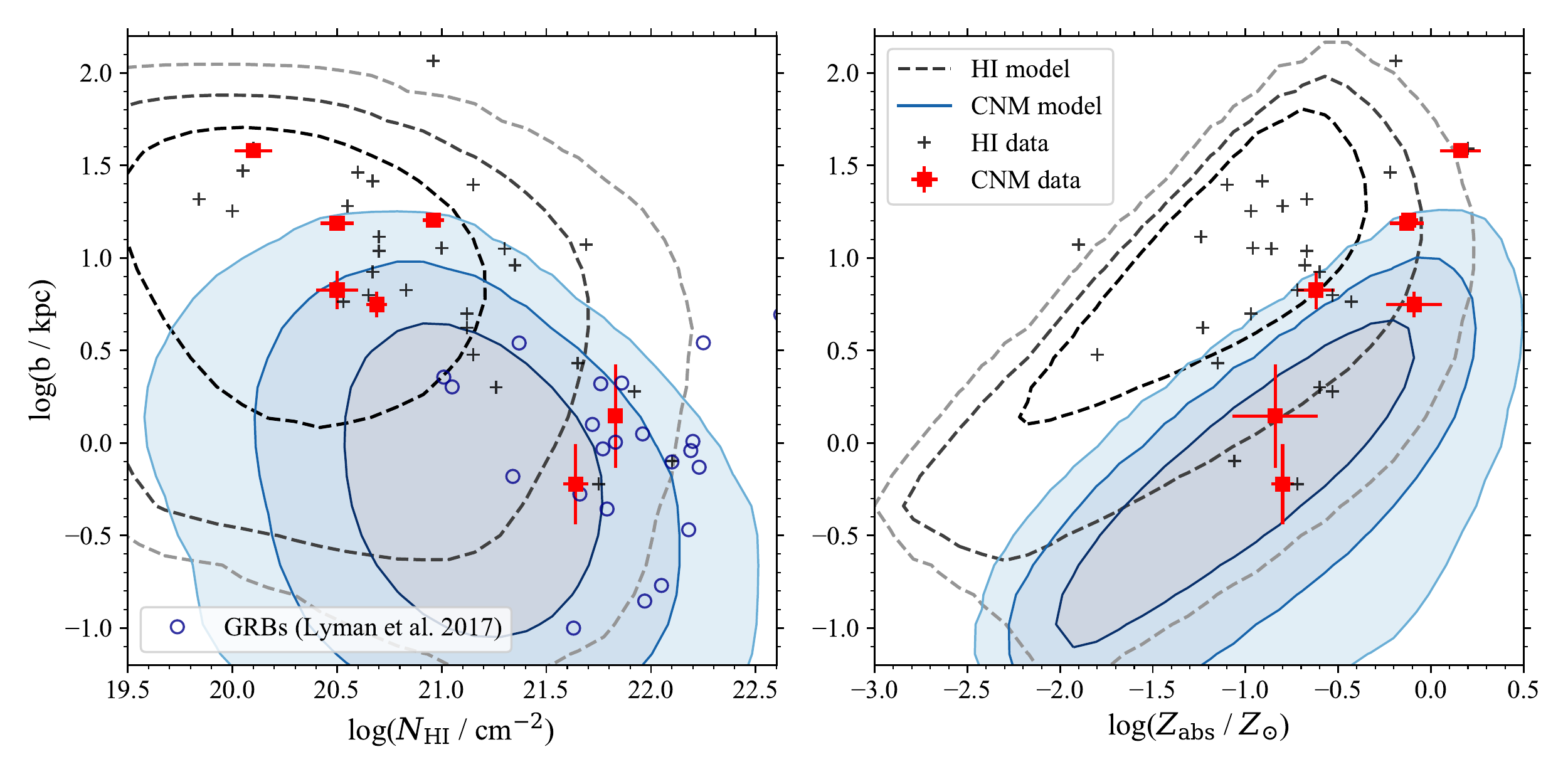}
    \caption{Model prediction for impact parameters as a function of \NHI\ and absorption metallicity. The blue-filled contours indicate the 99th, 95th, and 68th percentiles of the model distribution of CNM absorbers. The dashed contours in gray mark the percentiles of the model distribution of overall \HI\ absorbers. Red and black data points show CNM and \HI\ absorbers \citep{Moller2020}, respectively, with identified emission counterparts. The open circles in the left panel show data points for GRBs from \citet{Lyman2017}.
    The majority of the GRBs do not have metallicity measurements and are therefore not shown in the right panel.
    }
    \label{fig:b_Z_NHI}
\end{figure*}

    Lastly, in Fig.~\ref{fig:b_Z_NHI} we show the distribution of impact parameters as a function of \NHI\ and metallicity for the modeled CNM absorbers compared to the overall absorber population. In this figure, we show the seven CNM absorbers for which emission counterparts have been identified (see Table~\ref{tab:data}). For comparison purposes, we show the compilation of emission counterparts of \HI-selected absorbers by \citet{Moller2020}.
    As mentioned in the discussion by \citet{Krogager2018b}, the average projected radial extent of CNM gas at high redshift is expected to be roughly a factor of ten smaller than the typical extent of DLAs. This is in good agreement with our model where the median impact parameter of CNM absorbers is a factor 11 smaller than the median for DLAs.

\section{Discussion}
\label{discussion}

    An implicit assumption in our analysis is that the strong H$_2$ lines ($\NHH > 10^{18}$~cm$^{-2}$) analyzed by \citetalias{Balashev2018} are complete tracers of the CNM in our model. One independent way to probe the CNM gas in DLAs is through \HI\ 21-cm absorption lines. The detection rate of 21-cm absorption in DLAs at $z>2.2$ has been constrained to be $16^{+17}_{-9}$~\% \citep{Kanekar2014}. At face-value, this indicates that the CNM fraction of DLAs is higher than the 4~\% obtained from H$_2$ statistics and assuming CNM always bears self-shielded H$_2$ \citepalias{Balashev2018}. However, given the low-number statistics of high-redshift 21-cm absorbers, this fraction should be interpreted with great care. The CNM fraction derived from 21-cm absorbers depends strongly on the redshift criteria, for example, for $z>2.5$, which is more comparable to the sample analyzed by \citetalias{Balashev2018}, and the fraction drops to $\sim$7~\% (1/14; see also \citealt{Srianand2012}). The assumption that strong H$_2$ absorption systems trace the bulk of the CNM is therefore consistent with 21-cm absorption statistics.

    Absorption from \CI\ is a good tracer of CNM gas at high metallicity due to the increasing detectability with an increasing metal column and due to the more efficient cooling and dust-shielding from UV photons in high-metallicity gas. A comparison between our model prediction and the incidence derived for strong \CI\ absorption lines with $W_{\CI, 1560} > 0.4$~\AA\ by \citet{Ledoux2015} shows that $\sim$30~\% of the CNM gas at $\log (Z_{\rm abs}/Z_{\odot}) > -0.6$\footnote{This metallicity cut corresponds to the lowest metallicity observed in the sample by \citet{Ledoux2015}.} was identified by selecting based on strong \CI\ absorption.
    However, the simplifying assumptions in our model about the CNM covering factor and geometry affect the estimated fraction of the CNM identified using \CI-selection.
    
    We assume that the covering factor of CNM within \Rcnm\ is unity. While this may be an over-simplification, it is consistent with observations at lower redshifts \citep[e.g.,][]{Wiklind2018} and the observation of several \CI\ absorption components towards both lines of sight in the lensed quasar observed by \citet{Krogager2018b}.
    We also assume a spherical distribution of the cross section similar to \citetalias{Krogager2020}. While this may work well for the more extended warm gas phase dominating the strong \HI\ absorbers, the CNM on smaller scales may be distributed in a more flattened disk-like structure. Taking such a flattening into account would decrease the cross section of the CNM by roughly a factor of two. 
    Accounting for such a geometrical effect would bring the model expectation for \lCNM\ of metal-rich systems into closer agreement with the observed \lCI.  Nonetheless, some part of the cold, molecular gas may arise outside a flattened disk, for example, in accretion streams or material carried out in outflows. This may decrease the projected ellipticity on average, thereby lending more support to our spherical approximation.
    
    Taking our model at face value, we find that strong \CI-selected absorbers represent only a fraction of the total observable CNM (i.e., not obscured by dust). The inferred total {\lCNM} of $11.3\times10^{-3}$ yields a fraction of $13\pm5$~\% of the total CNM detectable through strong \CI\ absorption in current spectroscopic surveys of quasars.
    
    As alluded to above, a non-negligible part of the CNM has a dust obscuration that is too high in order to be identified in optical surveys. Based on our model, we find that the fractional completeness of \lCNM\ is $\sim$95~\% due to dust obscuration, giving a total unobscured CNM incidence of $l_{\rm \textsc{cnm} ,\, tot} \approx 12\times10^{-3}$. 

    We have so far only considered data from intervening quasar absorption systems. Another powerful probe of high-redshift galaxies is absorption systems observed in $\gamma$-ray burst afterglow spectra at the redshift of the host galaxy, the so-called GRB-DLAs. The GRB-DLAs probe very central regions of their host galaxies with impact parameters less than $\lesssim 5$~kpc on average \citep{Lyman2017}.
    For comparison purposes, in Fig.~\ref{fig:b_Z_NHI} we show a sample of GRB-DLAs. The small impact parameters overlap with the expected impact parameters of CNM sightlines from our model. This is consistent with the high fraction ($\gtrsim 25$~\%) of GRB-DLAs showing absorption from H$_2$ or \CI\ \citep{Bolmer2019, Heintz2019a}. The GRB sightlines, however, tend to probe higher \NHI, which may be a result of GRBs arising in regions of recent star formation where the gas column could be higher than the average galactic environment. A formal analysis of the GRB statistics in the context of our model is beyond the scope of this letter, yet we highlight that the similarities of ESDLAs \citep[$\logNHIcm > 21.7$;][]{Noterdaeme2014, Ranjan2020} and GRB-DLAs in terms of the high fraction of them showing signs of CNM and their small impact parameters are in agreement with our model.

\section{Summary}
\label{summary}

    We have presented a simple model to reproduce the statistics of high-redshift absorption systems featuring either H$_2$ or \CI\ absorption lines as tracers of the CNM.
    The aim of the model is to test the canonical two-phase model of the neutral ISM at high redshifts where the environment is significantly different from local galaxies (mainly in terms of metallicity and ambient ionizing flux). The main principle of the model relies on a pressure gradient throughout the galactic halo which determines at which point the neutral medium is able to support a stable CNM. The criterion for CNM to exist is determined by the minimum pressure, \Pmin, calculated following theoretical considerations by \citet{Wolfire2003}. This pressure-based prescription was incorporated into the model framework by \citetalias{Krogager2020} in order to predict the \HI\ and $Z_{\rm abs}$ of CNM absorption systems.
        
    We were able to constrain the single a priori unknown model parameter by matching the fraction of high-redshift ($z\approx2.5$) DLAs hosting CNM gas (as traced by H$_2$;  \citetalias{Balashev2018}). Our model then simultaneously reproduces the distribution functions of \NHI\ and \NHH. Furthermore, the distributions of metallicity, \NHI, and impact parameters are all in qualitative agreement with observations. However, due to the heterogeneous sample selection and limited statistics, it was not possible to formally quantify the goodness of fit.
        
    We find that the fraction of CNM identified by strong \CI\ absorption lines (with equivalent widths in excess of 0.4~\AA) amounts to $\sim$15~\% of the total CNM observable in optically-selected quasar surveys, and that the total incidence of CNM absorption is underestimated by $\sim$5~\% due to dust obscuration.
        
    Our model corroborates a simple picture for the neutral gas inside and around high-redshift galaxies. The mostly warm, neutral gas as probed by DLAs is extended over scales on the order of 10 kpc, whereas the cold neutral medium clouds are distributed on much smaller galactic scales, of the order 1~kpc, where the pressure is high enough.

\begin{acknowledgements}
    We thank R. Srianand for helpful discussions and comments on the manuscript. We thank Sergei Balashev for the very constructive review. The research leading to these results has received funding from the French {\sl Agence Nationale de la Recherche} under grant no ANR-17-CE31-0011-01 (project ``HIH2'' -- PI Noterdaeme).
\end{acknowledgements}

\bibliographystyle{aa}


\renewcommand{\arraystretch}{1.50}

\longtab{
\begin{longtable}{lccccccc}
\caption{Auxiliary data: Compilation of cold gas absorption systems. \label{tab:data}}\\
\hline\hline
Quasar & $z_{\rm abs}$ & $\log\left(\NHI\, /\, \textrm{\small cm}^{-2}\right)$ & $\log\left(Z\, /\, Z_{\odot}\right)$ & $\log\left({\rm N_{H_2}\, /\, \textrm{\small cm}}^{-2}\right)$ & $\log \left(P_{\rm th}\, /\, \textrm{\small K~cm}^{-3}\right)$ & $b / \textrm{\small kpc}$ & Reference   \\
\hline
\endfirsthead
\caption{continued.}\\
\hline\hline
Quasar & $z_{\rm abs}$ & $\log\left(\NHI\, /\, \textrm{\small cm}^{-2}\right)$ & $\log\left(Z\, /\, Z_{\odot}\right)$ & $\log\left({\rm N_{H_2}\, /\, \textrm{\small cm}}^{-2}\right)$ & $\log \left(P_{\rm th}\, /\, \textrm{\small K~cm}^{-3}\right)$ & $b / \textrm{\small kpc}$ & Reference   \\
\hline
\endhead
\hline
\endfoot
J0000+0048   & 2.526 & $20.80 \pm 0.10$ & \hphantom{$-$}$0.46 \pm 0.45$ & $20.43_{-0.02}^{+0.02}$ & $3.6 \pm 0.1$ & --             & (21)                 \\
Q0013$-$0029 & 1.973 & $20.83 \pm 0.05$ & $-0.59 \pm 0.05$              & $18.86_{-1.14}^{+1.14}$ & --            & --             & (23)                  \\
Q0027$-$1836 & 2.402 & $21.75 \pm 0.10$ & $-1.63 \pm 0.10$              & $17.30_{-0.07}^{+0.07}$ & --            & --             & (17)                \\
J0136+0440   & 2.779 & $20.73 \pm 0.01$ & $-0.58 \pm 0.03$              & $18.65_{-0.07}^{+0.07}$ & $4.1 \pm 0.1$ & --             & (4)                   \\
J0203+1134   & 3.387 & $21.26 \pm 0.08$ & $-1.25 \pm 0.10$              & $15.6 _{-0.8 }^{+0.8 }$ & --            & --             & (29, 5)     \\
J0216$-$0021 & 1.736 & $20.28 \pm 0.05$ & $-0.27 \pm 0.10$              & --                      & --            & --             & (15)                  \\
Q0347$-$3919 & 3.025 & $20.73 \pm 0.05$ & $-0.98 \pm 0.09$              & $14.55_{-0.09}^{+0.09}$ & --            & --             & (27)                   \\
Q0405$-$4418 & 2.595 & $20.90 \pm 0.10$ & $-1.02 \pm 0.12$              & $18.14_{-0.12}^{+0.07}$ & --            & --             & (27, 13)       \\
Q0551$-$3638 & 1.962 & $20.50 \pm 0.08$ & $-0.13 \pm 0.09$              & $17.42_{-0.93}^{+0.65}$ & --            & $15.4 \pm 1.0$ & (13, 10)        \\
J0643$-$5041 & 2.659 & $21.03 \pm 0.08$ & $-0.91 \pm 0.09$              & $18.54_{-0.01}^{+0.01}$ & $3.8 \pm 0.3$ & --             & (1)            \\
J0811+0838   & 1.905 & $20.10 \pm 0.10$ & $-0.11 \pm 0.17$              & --                      & --            & --             & (15)                  \\
J0812+3208   & 2.626 & $21.35 \pm 0.10$ & $-0.81 \pm 0.10$              & $19.89_{-0.03}^{+0.03}$ & $4.0 \pm 0.3$ & --             & (8)                  \\
J0815+2640   & 1.680 & $20.85 \pm 0.07$ & $-0.26 \pm 0.11$              & --                      & --            & --             & (15)                  \\
J0816+1446   & 3.287 & $22.00 \pm 0.10$ & $-1.10 \pm 0.10$              & $18.62_{-0.18}^{+0.18}$ & $3.7 \pm 0.1$ & --             & (7)                  \\
J0843+0221   & 2.786 & $21.82 \pm 0.11$ & $-1.52 \pm 0.09$              & $21.21_{-0.02}^{+0.02}$ & $4.9 \pm 0.1$ & --             & (3)                   \\
J0852+1935   & 1.787 & $20.00 \pm 0.20$ & \hphantom{$-$}$0.21 \pm 0.25$ & --                      & --            & --             & (15)                  \\
J0854+0317   & 1.566 & $20.68 \pm 0.05$ & $-0.30 \pm 0.08$              & --                      & --            & --             & (15)                  \\
J0857+1855   & 1.730 & $19.70 \pm 0.20$ & $-0.19 \pm 0.21$              & --                      & --            & --             & (15)                  \\
J0858+1749   & 2.625 & $20.40 \pm 0.01$ & $-0.63 \pm 0.02$              & $19.72_{-0.02}^{+0.02}$ & $3.8 \pm 0.1$ & --             & (4)                   \\
J0906+0548   & 2.567 & $20.13 \pm 0.01$ & $-0.18 \pm 0.07$              & $18.88_{-0.02}^{+0.02}$ & $4.7 \pm 0.1$ & --             & (4)                   \\
J0917+0154   & 2.107 & $21.00 \pm 0.07$ & $-0.18 \pm 0.12$              & $20.11_{-0.06}^{+0.06}$ & --            & --             & (15, 22)  \\
J0918+1636   & 2.580 & $20.96 \pm 0.05$ & $-0.12 \pm 0.05$              & $17.60_{-1.45}^{+1.45}$ & --            & $16.0 \pm 0.8$ & (6)                      \\
J0927+1543   & 1.731 & $21.00 \pm 0.20$ & $-0.16 \pm 0.21$              & --                      & --            & --             & (15)                  \\
J0946+1216   & 2.607 & $21.15 \pm 0.02$ & $-0.48 \pm 0.01$              & $19.97_{-0.02}^{+0.02}$ & $4.4 \pm 0.1$ & --             & (4)                   \\
J1037$-$2703 & 2.139 & $19.70 \pm 0.10$ & $-0.26 \pm 0.11$              & --                      & --            & --             & (13, 27)       \\
J1047+2057   & 1.775 & $20.58 \pm 0.05$ & $-0.18 \pm 0.12$              & --                      & --            & --             & (15)                  \\
J1117+1437   & 2.001 & $19.80 \pm 0.10$ & \hphantom{$-$}$0.35 \pm 0.14$ & $18.0 _{-1.0 }^{+1.0 }$ & --            & --             & (15, 22)  \\
J1122+1437   & 1.554 & $20.18 \pm 0.10$ & $-0.62 \pm 0.15$              & --                      & --            & --             & (15)                  \\
J1133$-$0057 & 1.705 & $21.00 \pm 0.30$ & $-0.44 \pm 0.31$              & --                      & --            & --             & (15)                  \\
J1143+1420   & 2.323 & $21.64 \pm 0.06$ & $-0.80 \pm 0.06$              & $18.30_{-0.10}^{+0.10}$ & --            & $0.6 \pm 0.3$  & (25)                     \\
J1146+0743   & 2.840 & $21.54 \pm 0.01$ & $-0.57 \pm 0.02$              & $18.82_{-0.03}^{+0.03}$ & $4.6 \pm 0.2$ & --             & (4)                   \\
Q1232+0815   & 2.338 & $20.90 \pm 0.09$ & $-1.32 \pm 0.12$              & $19.57_{-0.11}^{+0.11}$ & $3.8 \pm 0.1$ & --             & (2)                   \\
J1236+0010   & 3.033 & $20.78 \pm 0.01$ & $-0.58 \pm 0.04$              & $19.76_{-0.01}^{+0.01}$ & $3.6 \pm 0.6$ & --             & (4)                   \\
J1237+0647   & 2.691 & $20.00 \pm 0.15$ & \hphantom{$-$}$0.34 \pm 0.12$ & $19.21_{-0.13}^{+0.13}$ & $3.7 \pm 0.2$ & --             & (19)                \\
J1248+2848   & 1.513 & $20.35 \pm 0.15$ & $-0.04 \pm 0.17$              & --                      & --            & --             & (15)                  \\
J1302+2111   & 1.656 & $21.07 \pm 0.07$ & $-0.59 \pm 0.10$              & --                      & --            & --             & (15)                  \\
J1306+2815   & 2.012 & $19.70 \pm 0.20$ & \hphantom{$-$}$0.24 \pm 0.22$ & --                      & --            & --             & (15)                  \\
J1311+2225   & 3.092 & $20.75 \pm 0.10$ & $-0.61 \pm 0.14$              & $19.69_{-0.01}^{+0.01}$ & --            & --             & (15, 22)  \\
J1314+0543   & 1.583 & $20.07 \pm 0.10$ & \hphantom{$-$}$0.24 \pm 0.12$ & --                      & --            & --             & (15)                  \\
Q1331+0170   & 1.777 & $21.17 \pm 0.07$ & $-1.22 \pm 0.04$              & $19.71_{-0.07}^{+0.07}$ & $3.7 \pm 0.3$ & --             & (9)                  \\
J1337+3152   & 3.174 & $21.36 \pm 0.10$ & $-1.45 \pm 0.22$              & $14.09_{-0.04}^{+0.04}$ & $3.1 \pm 0.3$ & --             & (28)                   \\
J1439+1117   & 2.418 & $20.10 \pm 0.09$ & \hphantom{$-$}$0.16 \pm 0.11$ & $19.38_{-0.10}^{+0.10}$ & --            & $38$\tablefootmark{a} & (18, 26)     \\
J1441+2737   & 4.224 & $20.95 \pm 0.10$ & $-0.63 \pm 0.10$              & $18.29_{-0.08}^{+0.07}$ & --            & --             & (14)           \\
Q1444+0126   & 2.087 & $20.25 \pm 0.07$ & $-0.80 \pm 0.09$              & $18.15_{-0.15}^{+0.15}$ & $4.1 \pm 0.3$ & --             & (27)                   \\
J1456+1609   & 3.350 & $21.70 \pm 0.10$ & $-1.32 \pm 0.11$              & $17.10_{-0.09}^{+0.09}$ & --            & --             & (20)                \\
J1513+0352   & 2.464 & $21.83 \pm 0.02$ & $-0.84 \pm 0.23$              & $21.31_{-0.01}^{+0.01}$ & $4.4 \pm 0.1$ & $1.4 \pm 0.9$  & (24)                     \\
J1615+2648   & 2.118 & $20.55 \pm 0.10$ & $-0.05 \pm 0.15$              & --                      & --            & --             & (15)                  \\
J1623+1355   & 1.751 & $20.10 \pm 0.10$ & $-0.01 \pm 0.26$              & --                      & --            & --             & (15)                  \\
J1646+2329   & 1.998 & $19.75 \pm 0.10$ & \hphantom{$-$}$0.13 \pm 0.18$ & $18.02_{-0.11}^{+0.11}$ & --            & --             & (15, 22)  \\
J1705+3543   & 2.038 & $20.62 \pm 0.12$ & \hphantom{$-$}$0.07 \pm 0.14$ & --                      & --            & --             & (15)                  \\
J2100$-$0641 & 3.091 & $21.05 \pm 0.15$ & $-0.73 \pm 0.15$              & $18.76_{-0.03}^{+0.03}$ & $3.5 \pm 1.2$ & --             & (9)                  \\
J2123$-$0050 & 2.060 & $19.18 \pm 0.15$ & \hphantom{$-$}$0.12 \pm 0.15$ & $17.94_{-0.01}^{+0.01}$ & --            & --             & (15, 22)  \\
J2140$-$0321 & 2.340 & $22.40 \pm 0.10$ & $-1.05 \pm 0.13$              & $20.13_{-0.07}^{+0.07}$ & $4.6 \pm 0.3$ & --             & (20)                \\
J2225+0527   & 2.131 & $20.69 \pm 0.05$ & $-0.09 \pm 0.15$              & $19.40_{-0.10}^{+0.10}$ & --            & $5.6 \pm 0.9$  & (11, 10)     \\
J2229+1414   & 1.585 & $19.80 \pm 0.15$ & \hphantom{$-$}$0.04 \pm 0.25$ & --                      & --            & --             & (15)                  \\
J2232+1242   & 2.230 & $21.75 \pm 0.03$ & $-1.48 \pm 0.05$              & $18.56_{-0.02}^{+0.02}$ & --            & --             & (25)                     \\
Q2231$-$0015 & 2.066 & $20.56 \pm 0.10$ & $-0.74 \pm 0.16$              & --                      & $3.9 \pm 0.5$ & --             & (9)                  \\
J2257$-$1001 & 1.836 & $20.40 \pm 0.07$ & \hphantom{$-$}$0.03 \pm 0.18$ & $19.50_{-0.10}^{+0.10}$ & --            & --             & (15, 22)  \\
Q2318$-$1107 & 1.989 & $20.68 \pm 0.05$ & $-0.85 \pm 0.06$              & $15.49_{-0.03}^{+0.03}$ & --            & --             & (17)                \\
J2331$-$0908 & 2.143 & $21.15 \pm 0.15$ & $-0.54 \pm 0.15$              & $20.57_{-0.05}^{+0.05}$ & --            & --             & (15, 22)  \\
J2334$-$0908 & 2.287 & $20.50 \pm 0.07$ & $-1.33 \pm 0.11$              & --                      & $3.7 \pm 0.2$ & --             & (27)                   \\
J2336$-$1058 & 1.829 & $20.33 \pm 0.02$ & $-0.22 \pm 0.10$              & $19.00_{-0.12}^{+0.12}$ & --            & --             & (15, 22)  \\
J2340$-$0053 & 2.054 & $20.35 \pm 0.15$ & $-0.92 \pm 0.03$              & $18.47_{-0.04}^{+0.04}$ & $3.9 \pm 0.4$ & --             & (9, 22)  \\
Q2343+1232   & 2.431 & $20.40 \pm 0.07$ & $-0.87 \pm 0.10$              & $13.69_{-0.09}^{+0.09}$ & --            & --             & (17)                \\
J2347$-$0051 & 2.588 & $20.47 \pm 0.01$ & $-0.82 \pm 0.04$              & $19.44_{-0.01}^{+0.01}$ & $3.9 \pm 0.5$ & --             & (4)                   \\
J2350$-$0052 & 2.426 & $20.50 \pm 0.10$ & $-0.62 \pm 0.10$              & $18.52_{-0.49}^{+0.29}$ & $4.3 \pm 0.1$ & $6.7 \pm 1.6$  & (16, 12)  \\
\hline

\end{longtable}
\tablefoot{
\tablefoottext{a}{No uncertainty was reported by \citet{Rudie2017}.}
}
\tablebib{(1) \citet{AlbornozVasquez2014}; (2) \citet{Balashev2011}; (3) \citet{Balashev2017}; (4) \citet{Balashev2019}; (5) \citet{Ellison2001a}; (6) \citet{Fynbo2011}; (7) \citet{Guimaraes2012}; (8) \citet{Jorgenson2009}; (9) \citet{Jorgenson2010}; (10) \citet{Kanekar2020}; (11) \citet{Krogager2016a}; (12) \citet{Krogager2017}; (13) \citet{Ledoux2003}; (14) \citet{Ledoux2006_molecules}; (15) Ledoux et al. in preparation; (16) \citet{Noterdaeme2007a}; (17) \citet{Noterdaeme2007b}; (18) \citet{Noterdaeme2008b}; (19) \citet{Noterdaeme2010b}; (20) \citet{Noterdaeme2015a}; (21) \citet{Noterdaeme2017}; (22) \citet{Noterdaeme2018}; (23) \citet{Petitjean2002}; (24) \citet{Ranjan2018}; (25) \citet{Ranjan2020}; (26) \citet{Rudie2017}; (27) \citet{Srianand2005}; (28) \citet{Srianand2010}; (29) \citet{Srianand2012}
}
}

\end{document}